\documentclass[twoside,11pt]{article} %%% Please do not change this class !!! %%%

\usepackage{fancyhdr}    % headings style
\usepackage{epic}
\usepackage{eepic}
\usepackage{epsf}
\usepackage{graphicx}
\usepackage{rotating}
\begin{document}
 \baselineskip=11pt

\title{The Bargmann-Wigner Formalism for Higher Spins (up to 2)}

%\hspace{.25mm}

\author{{\bf Valeriy V. Dvoeglazov}\\
Unidad Acad\'emica de F\'{\i}sica\\
Universidad de Zacatecas, 
Apartado Postal 636, Suc. 3 Cruces\\
Zacatecas 98064, Zac., M\'exico\\
URL: http://planck.reduaz.mx/\~{}valeri/ \\
e-mail address: valeri@planck.reduaz.mx} 

\vspace{2mm}

\date{}

\maketitle

\begin{abstract}
On the basis of our recent modifications of the Dirac formalism we
generalize the Bargmann-Wigner formalism for higher spins to be compatible
with other formalisms for bosons. Relations with dual electrodynamics, with
the Ogievetskii-Polubarinov notoph and the Weinberg 2(2S+1) theory are
found.  Next, we proceed to derive the equations for the symmetric tensor of the second rank on the basis of the Bargmann-Wigner formalism in a straightforward way. The symmetric multispinor of the fourth rank is used. It is constructed out of the Dirac 4-spinors. Due to serious problems with the interpretation of the results obtained on using the standard procedure we generalize it and obtain the spin-2 relativistic equations, which are consistent with the previous one. 
We introduce the dual analogues of the Riemann tensor and
derive corresponding dynamical equations in the Minkowski space. Relations
with the Marques-Spehler chiral gravity theory are discussed.
The importance of the 4-vector field (and its gauge part) is pointed out.
\end{abstract}

%\clearpage % delate this line

{\Large{

%\vspace{2cm} % delate this line

\section{Introduction.}

Recent advances in astrophysics~\cite{scalarmatos} suggest the existence of fundamental scalar 
fields~\cite{Hua,Hua1}. It can be used in the consideration of the gravitational phenomena beyond the frameworks
of the general relativity (for instance, in some models of quantum gravity). 
On the other hand, the $(1/2,1/2)$ representation of the Lorentz group provides 
suitable frameworks for introduction of the $S=0$ field, Ref.~\cite{Weinb}. 
In this paper, starting from the very beginning we propose a generalized theory in 
the 4-vector representation, for the antisymetric tensor field of the second rank as well. 
The results can be useful in any theory dealing with the light phenomena. 

The general scheme for derivation of higher-spin equations
was given in~\cite{BW}. A field of rest mass $m$ and spin $s \geq {1\over
2}$ is represented by a completely symmetric multispinor of rank $2s$.
The particular cases $s=1$ and $s={3\over 2}$ have been given in the
textbooks, e.~g., ref.~\cite{Lurie}. The spin-2 case can also be of some
interest because it is generally believed that the essential features of
the gravitational field are  obtained from transverse components of the
$(2,0)\oplus (0,2)$  representation of the Lorentz group. Nevertheless,
questions of the redundant components of the higher-spin relativistic
equations are not yet understood in detail~\cite{Kirch}.

The plan of my talk is following:

\begin{itemize}

\item
Antecedents. Motivations. The mapping between the Weinberg-Tucker-Hammer (WTH) formulation
and antisymmetric tensor (AST) fields of the 2nd rank. 

\item
The Modified Bargmann-Wigner (BW) formalism.  Pseudovector potential. Parity.

\item
The matrix form of the general equation in the $(1/2,1/2)$ representation.
Lagrangian in the matrix form. Masses.

\item
The Standard Basis and the Helicity Basis.

\item
Dynamical invariants. Field operators. Propagators.
The indefinite metric.

\item
The Spin-2 Framework. The dual Riemann tensors. Equations for the symmetric tensor of the 2nd rank.

\end{itemize}

\section{Preliminaries.}

I am going to give an overview of my previous works in order you 
to understand motivations better. In Ref.~\cite{Hua,Hua1}
I derived the Maxwell-like equations with the additional gradient of a scalar field ${\chi}$ from the first principles. 
Here they are:
\begin{eqnarray}
&&{\nabla}\times {\bf
E}=-\frac{1}{c}\frac{\partial {\bf B}}{\partial t} + {\nabla} {\it Im} \chi \,, \label{1}\\
&&{\nabla }\times {\bf B}=\frac{1}{c}\frac{\partial {\bf
E}}{\partial t}  +{\nabla} {\it Re} \chi\,,\label{2}\\
&&{\nabla}\cdot {\bf E}=-\frac{1}{c} \frac{\partial}{\partial
t} {\it Re}\chi \,,\label{3}\\
&&{\nabla }\cdot {\bf B}= \frac{1}{c} \frac{\partial}{\partial t} {\it Im} \chi \,.  \label{4}
\end{eqnarray}
The $\chi$ may depend on the ${\bf E}, {\bf B}$ fields, so we can have the non-linear electrodynamics.
Of course, similar equations can be obtained 
in the massive case $m\neq 0$, i.e., within the Proca-like theory.
We should then consider
\begin{equation}
(E^2 -c^2 {\bf p}^2 - m^2 c^4 ) \Psi^{(3)} =0\, .\label{5}
\end{equation}
In the spin-1/2 case the analogous equation  can be written 
for the two-component spinor ($c=\hbar =1$)
\begin{equation}
(E I^{(2)} - {\sigma}\cdot {\bf p})
(E I^{(2)} + {\sigma}\cdot {\bf p})\Psi^{(2)} = m^2 \Psi^{(2)}\,,
\end{equation}
or, in the 4-component form\footnote{There exist various generalizations
of the Dirac formalism. For instance, the Barut generalization
is based on
\begin{equation}
[i\gamma_\mu \partial_\mu +  a (\partial_\mu \partial_\mu)/m - \kappa ] \Psi =0\,,
\end{equation}
which can describe states of different masses. If one fixes the parameter $a$ by the requirement that the equation gives the state with the classical anomalous magnetic moment, then $m_2 =
m_1 (1+\frac{3}{2\alpha})$, i.e., it gives the muon mass. Of course, one can propose a generalized equation:
\begin{equation}
[i\gamma_\mu \partial_\mu +  a +b \partial_\mu \partial_\mu + \gamma_5 (c+d \partial_\mu \partial_\mu ) ] \Psi =0
\,;
\end{equation}
and, perhaps, even that of higher orders in derivatives.}
\begin{equation}
[i\gamma_\mu \partial_\mu +m_1 +m_2 \gamma^5 ] \Psi^{(4)} = 0\,.
\end{equation}
In the spin-1 case  we have
\begin{equation}
(E I^{(3)} - {\bf S}\cdot {\bf p})
(E I^{(3)} + {\bf S}\cdot {\bf p}){\Psi}^{(3)} 
- {\bf p} ({\bf p}\cdot {\Psi}^{(3)})= m^2 \Psi^{(3)}\,,
\end{equation}
that lead to (\ref{1}-\ref{4}), when $m=0$. We can continue writing down 
equations for higher spins in a similar fashion.

On this basis we are ready to generalize the BW formalism~\cite{BW,Lurie}. Why is that convenient? In Ref.~\cite{dv-hpa,wig} I presented the mapping between the WTH equation, Ref.~\cite{WTH,WTH1}, and the equations for AST fields. The equation for a 6-component field function is\footnote{ In order to have solutions satisfying the Einstein dispersion relations $E^2 -{\bf p}^2 =m^2$ we have to assume $B/(A+1)= 1$, or $B/(A-1)=1$.}
\begin{equation}
[\gamma_{\alpha\beta} p_\alpha p_\beta +A p_\alpha p_\alpha +Bm^2 ]
\Psi^{(6)} =0\,.
\end{equation}
Corresponding equations for the AST fields are:
\begin{eqnarray}
&&\partial_\alpha\partial_\mu F_{\mu\beta}^{(1)}
-\partial_\beta\partial_\mu F_{\mu\alpha}^{(I1)}
+ \frac{A-1}{2} \partial_\mu \partial_\mu F_{\alpha\beta}^{(1)}
-\frac{B}{2} m^2 F_{\alpha\beta}^{(1)} = 0\nonumber\\
\label{wth1}\\
&&\partial_\alpha\partial_\mu F_{\mu\beta}^{(2)}
-\partial_\beta\partial_\mu F_{\mu\alpha}^{(2)}
- \frac{A+1}{2} \partial_\mu \partial_\mu F_{\alpha\beta}^{(2)}
+\frac{B}{2} m^2 F_{\alpha\beta}^{(2)}=0\nonumber\\
\label{wth2}
\end{eqnarray}
depending on the parity properties of $\Psi^{(6)}$ (the first case corresponds 
to the eigenvalue $P=-1$; the second one, to $P=+1$). 

We noted:

\begin{itemize}

\item
One can derive the equations for the dual tensor $\tilde F_{\alpha\beta}$,
which are similar to (\ref{wth1},\ref{wth2}), Ref.~\cite{dv-cl,dv-hpa}.

\item
In the Tucker-Hammer case ($A=1$, $B=2$), the first equation gives the Proca theory $\partial_\alpha \partial_\mu 
F_{\mu\beta} -
\partial_\beta \partial_\mu F_{\mu\alpha} = m^2 F_{\alpha\beta}$. In the second case one finds something different, $\partial_\alpha \partial_\mu F_{\mu\beta} -
\partial_\beta \partial_\mu F_{\mu\alpha} = (\partial_\mu \partial_\mu - m^2 ) F_{\alpha\beta}$

\item
If $\Psi^{(6)}$ has no definite parity, e.~g., $\Psi^{(6)} = 
\mbox{column} ({\bf E}+i{\bf B}\,\,\, {\bf B}+i{\bf E}\, )$, the equation
for the AST field will contain both the tensor and the dual tensor:
\begin{equation}
\partial_\alpha \partial_\mu F_{\mu\beta}
-\partial_\beta \partial_\mu F_{\mu\alpha}
=\frac{1}{2} \partial^2 F_{\alpha\beta} +
[-\frac{A}{2} \partial^2 + \frac{B}{2} m^2] \tilde
F_{\alpha\beta}.\label{pv1}
\end{equation}

\item
Depending on the relation between $A$ and $B$ and on which
parity solution  do we consider, the WTH equations may describe different mass states. For instance, when $A=7$ and $B=8$ we have the second mass state $(m^{\prime})^2 = 4m^2/3$.

\end{itemize}

We tried to find relations between the generalized WTH theory
and other spin-1 formalisms.  Therefore, we were forced to modify the Bargmann-Wigner formalism~\cite{dv-cl,dv-ps}. 

The Bargmann-Wigner formalism
for constructing  of high-spin particles has been given
in~\cite{BW,Lurie}. However, they claimed
explicitly that they constructed $(2S+1)$ states (the
Weinberg-Tucker-Hammer theory has  essentially $2(2S+1)$
components).  The standard Bargmann-Wigner formalism for $S=1$ is based on
the following set 
\begin{eqnarray} \left [ i\gamma_\mu \partial_\mu +m \right
]_{\alpha\beta} \Psi_{\beta\gamma} &=& 0\,,\label{bw1s}\\ \left [
i\gamma_\mu \partial_\mu +m \right ]_{\gamma\beta} \Psi_{\alpha\beta} &=&
0\,, \label{bw2s} \end{eqnarray} 
If one has
\begin{equation} \Psi_{\left \{ \alpha\beta
\right \} } = (\gamma_\mu R)_{\alpha\beta} A_\mu +
(\sigma_{\mu\nu} R)_{\alpha\beta} F_{\mu\nu}\,,
\end{equation} with
\begin{equation} 
R = e^{i\varphi}
\pmatrix{\Theta&0\cr 0&-\Theta\cr}\,\quad
\Theta=\pmatrix{0&-1\cr
1&0\cr}
\end{equation} in the spinorial
representation of $\gamma$-matrices we obtain
the Duffin-Proca-Kemmer equations:
\begin{eqnarray}
&&\partial_\alpha F_{\alpha\mu} = {m\over 2} A_\mu\,,\\
&& 2m F_{\mu\nu} = \partial_\mu A_\nu - \partial_\nu A_\mu\,.
\end{eqnarray}
After the corresponding
re-normalization $A_\mu \rightarrow 2m A_\mu$, we obtain the standard
textbook set:
\begin{eqnarray} &&\partial_\alpha
F_{\alpha\mu} = m^2 A_\mu\,,\\ && F_{\mu\nu} = \partial_\mu A_\nu -
\partial_\nu A_\mu\,.  \end{eqnarray} 
It gives the Tucker-Hammer equation for the antisymmetric tensor field. 
How can one obtain other equations
following the Weinberg-Tucker-Hammer approach?
The third equation~\cite{wig}  can be obtained in a  simple way: use, instead of
$(\sigma_{\mu\nu} R) F_{\mu\nu}$, another symmetric matrix $(\gamma^5
\sigma_{\mu\nu} R)  F_{\mu\nu}$, see~\cite{dv-ps}.
And what about the second and the fourth equations?  I suggest:

\begin{itemize}

\item
to use, see above and~\cite{g1}:
\begin{equation}
[i\gamma_\mu \partial_\mu +m ] \Psi =0 \Rightarrow
[i\gamma_\mu \partial_\mu +m_1 +m_2\gamma_5 ] \Psi =0\,;
\end{equation}

\item
to use the Barut extension:
\begin{equation}
[i\gamma_\mu \partial_\mu +m ] \Psi =0 \Rightarrow
[i\gamma_\mu \partial_\mu +a{\partial_\mu \partial_\mu\over m} +\kappa ] \Psi
=0\,.
\end{equation}

\item
Next, we can introduce the sign operator in the Dirac equations which are the input for the formalism for symmetric 2-rank spinor:
\begin{eqnarray}
\left [ i\gamma_\mu \partial_\mu + \epsilon_1 m_1 +\epsilon_2 m_2 \gamma_5
\right ]_{\alpha\beta} \Psi_{\beta\gamma} &=&0\,,\\
\left [ i\gamma_\mu
\partial_\mu + \epsilon_3 m_1 +\epsilon_4 m_2 \gamma_5 \right ]_{\gamma\beta}
\Psi_{\alpha\beta} &=&0\,,
\end{eqnarray}
In such a way we can enlarge the set of possible states.

\end{itemize}
We begin with
\begin{eqnarray}
\left [ i\gamma_\mu \partial_\mu + a -b (\partial_\mu \partial_\mu) + \gamma_5 (c- d(\partial_\mu \partial_\mu) )
\right ]_{\alpha\beta} \Psi_{\beta\gamma} &=&0\,,\\
\left [ i\gamma_\mu
\partial_\mu + a -b (\partial_\mu \partial_\mu) - \gamma_5 (c- d(\partial_\mu \partial_\mu) ) \right ]_{\alpha\beta}
\Psi_{\gamma\beta} &=&0\,,
\end{eqnarray}
$(\partial_\mu \partial_\mu)$ is the d'Alembertian.
Thus, we obtain the Proca-like equations:
\begin{eqnarray} &&\partial_\nu A_\lambda - \partial_\lambda A_\nu - 2(a
+b \partial_\mu \partial_\mu ) F_{\nu \lambda} =0\,,\\ &&\partial_\mu
F_{\mu \lambda} = {1\over 2} (a +b \partial_\mu \partial_\mu) A_\lambda +
{1\over 2} (c+ d \partial_\mu \partial_\mu) \tilde A_\lambda\,,
\end{eqnarray}
$\tilde A_\lambda$ is the axial-vector potential (analogous to that
used in the Duffin-Kemmer set for $J=0$). Additional constraints are:
\begin{eqnarray}
&&i\partial_\lambda A_\lambda + ( c+d\partial_\mu \partial_\mu) \tilde \phi
=0\,,\\
&&\epsilon_{\mu\lambda\kappa \tau} \partial_\mu F_{\lambda\kappa } =0\,,
( c+ d \partial_\mu \partial_\mu ) \phi =0\,.
\end{eqnarray}

The spin-0 Duffin-Kemmer equations are:
\begin{eqnarray}
&&(a+b \partial_\mu \partial_\mu) \phi = 0\,, 
i\partial_\mu \tilde A_\mu  - (a+b\partial_\mu \partial_\mu) \tilde
\phi =0\,,\\
&&(a+b\, \partial_\mu \partial_\mu) \tilde A_\nu + (c+d\,\partial_\mu
\partial_\mu) A_\nu + i (\partial_\nu \tilde \phi) =0\,.
\end{eqnarray}
The additional constraints are:
\begin{equation}
\partial_\mu \phi =0\,,
\partial_\nu \tilde A_\lambda - \partial_\lambda \tilde
A_\nu +2 (c+d\partial_\mu \partial_\mu ) F_{\nu \lambda} = 0\,.
\end{equation}
In such a way the spin states are {\it mixed} through the 4-vector potentials.
After elimination of the 4-vector potentials we obtain
the equation for the AST field of the second rank:
\begin{eqnarray}
&&\left [ \partial_\mu \partial_\nu F_{\nu\lambda} - \partial_\lambda
\partial_\nu F_{\nu\mu}\right ]   + \left [ (c^2 - a^2) - 2(ab-cd)
\partial_\mu\partial_\mu  + \right.\nonumber\\
&&\left. +(d^2 -b^2)
(\partial_\mu\partial_\mu)^2 \right ] F_{\mu\lambda} = 0\,,
\end{eqnarray}
which should be compared with our
previous equations which follow from the Weinberg-like formulation.
Just put:
\begin{eqnarray}
c^2 - a^2 \Rightarrow {-Bm^2 \over 2}\,,&\qquad& c^2 - a^2 \Rightarrow
+{Bm^2 \over 2}\,,\\
-2(ab-cd) \Rightarrow {A-1\over 2}\,,&\qquad&
+2(ab-cd) \Rightarrow {A+1\over 2}\,,\\
b=\pm d\,.&\qquad&
\end{eqnarray}

In the case with the sign operators we have 16 possible combinations, but 4 of them give the same
sets of the Proca-like equations. We obtain~\cite{dv-cl}:
\begin{eqnarray} 
&&\partial_\mu A_\lambda - \partial_\lambda A_\mu + 2m_1 A_1 F_{\mu \lambda}
+im_2 A_2 \epsilon_{\alpha\beta\mu\lambda} F_{\alpha\beta} =0\,,\\
&&\partial_\lambda
F_{\mu \lambda} - \frac{m_1}{2} A_1 A_\mu -\frac{m_2}{2} B_2 \tilde
A_\mu=0\,,
\end{eqnarray} 
with 
$A_1 = (\epsilon_1 +\epsilon_3) /2$,
$A_2 = (\epsilon_2 +\epsilon_4 )/ 2$,
$B_1 = (\epsilon_1 -\epsilon_3 )/ 2$,
and
$B_2 = (\epsilon_2 -\epsilon_4 )/ 2$. See the additional constraints in the cited paper~\cite{dv-cl}.
So, we have the dual tensor and the pseudovector potential
in the Proca-like sets. The pseudovector potential is the same as that
which enters in the Duffin-Kemmer set for the spin 0. 

Keeping in mind these observations, permit us
to repeat the derivation procedure for the Proca equations
from the equations of Bargmann and Wigner for a  totally
{\it symmetric} spinor of the second rank in a different way. 
As opposed to the previous consideration 
one can put
\begin{equation}
\Psi_{\{\alpha\beta\}} = (\gamma^\mu R)_{\alpha\beta} (c_a m A_\mu + c_f
F_\mu) +c_A m (\gamma^5 \sigma^{\mu\nu} R)_{\alpha\beta}
A_{\mu\nu} + c_F (\sigma^{\mu\nu} R)_{\alpha\beta}
F_{\mu\nu}\, .\label{si} \end{equation}
$\gamma^5$ is assumed to be diagonal.  The constants $c_i$ are some
numerical dimensionless coefficients. The reflection operator
$R$ has the following properties:
\begin{eqnarray}
&& R^T = -R\,,\quad R^\dagger =R = R^{-1}\,,\quad
R^{-1} \gamma^5 R = (\gamma^5)^T\,,\\
&& R^{-1}\gamma^\mu R = -(\gamma^\mu)^T\,,\quad
R^{-1} \sigma^{\mu\nu} R = - (\sigma^{\mu\nu})^T\,.
\end{eqnarray}
that are necessary for the expansion (\ref{si})
to be possible in such a form, i.~e., $\gamma^{\mu} R$, $\sigma^{\mu\nu} R$
$\gamma^5\sigma^{\mu\nu} R$ are assumed to be  {\it symmetric}
matrices.
The substitution of the preceding expansion into the
Bargmann-Wigner set~\cite{Lurie}
\begin{eqnarray}
\left [ i\gamma^\mu
\partial_\mu -m \right ]_{\alpha\beta} \Psi_{\{\beta\gamma\}} (x) &=&
0\,,\label{bw1}\\
\left [ i\gamma^\mu \partial_\mu -m \right
]_{\gamma\beta} \Psi_{\{\alpha\beta\}} (x) &=& 0\,  \label{bw2}
\end{eqnarray}
gives us the new equations of Proca:
\begin{eqnarray}
&& c_a m (\partial_\mu A_\nu - \partial_\nu A_\mu ) +
c_f (\partial_\mu F_\nu -\partial_\nu F_\mu ) =\nonumber\\
&& =ic_A m^2 \epsilon_{\alpha\beta\mu\nu} A^{\alpha\beta} +
2 m c_F F_{\mu\nu}\,,\label{pr1} \\
&& c_a m^2 A_\mu + c_f m F_\mu =
i c_A m \epsilon_{\mu\nu\alpha\beta} \partial^\nu A^{\alpha\beta} +
2 c_F \partial^\nu F_{\mu\nu}\, . \label{pr2}
\end{eqnarray}
In the case $c_a=1$, $c_F={1\over 2}$, $c_f=c_A=0$
they reduce to the ordinary Proca equations.\footnote{However, we noticed
that the division by $m$ in the first equation is {\it not} well-defined
operation in the case when somebody becomes interested
in the  $m\rightarrow 0$ limit later on.  Probably, in order
to avoid this dark point, one may wish to write the Dirac equation
in the form
$\left [ (i\gamma^\mu \partial_\mu)/ m-1\right ]\psi (x)= 0$,
the one that follows immediately in the
derivation of the  Dirac equation on the basis of the Ryder
relation~\cite{Ryder} and the Wigner rules
for a {\it boost} of the field function from the
system with zero linear momentum.} In the generalized case
one obtains dynamical equations that connect the
photon, the {\it notoph} and their potentials.  Divergent parts
(in $m\rightarrow 0$) of field functions and of
dynamical variables should be removed by corresponding `gauge'
transformations  (either electrtomagnetic gauge transformations or
Kalb-Ramond gauge transformations).  It is well
known that the massless {\it notoph} field  turns out to be
a pure longitudinal field when one
keeps in mind $\partial_\mu A^{\mu\nu}= 0$.

Apart from these dynamical equations, we can obtain a set of
constraints by means of subtraction of the equations of
Bargmann and Wigner (instead of their addition as in (\ref{pr1},\ref{pr2})).
They are read
\begin{equation} mc_a
\partial^\mu A_\mu + c_f \partial^\mu F_\mu =0,\, \quad\mbox{and} \quad
mc_A \partial^\alpha A_{\alpha\mu} + {i\over 2} c_F
\epsilon_{\alpha\beta\nu\mu} \partial^\alpha F^{\beta\nu} = 0\, .
\end{equation}
that suggest
$\widetilde F^{\mu\nu} \sim im A^{\mu\nu}$ $ F^\mu \sim m
A^\mu$, like in~\cite{Og}.

Following~\cite[Eqs.(9,10)]{Og}
we proceed in the construction of the ``potentials" for the
{\it notoph}: $A_{\mu\nu} ({\bf p})\sim\left [\epsilon_\mu^{(1)} ({\bf p})
\epsilon_\nu^{(2)} ({\bf p})- \epsilon_\nu^{(1)} ({\bf
p})\epsilon_\mu^{(2)} ({\bf p})\right ]$ upon using explicit forms of the
polarization vectors in the linear momentum space (e.~g.,
ref.~\cite{Weinb}).  One obtains
\begin{eqnarray} 
A^{\mu\nu} ({\bf p}) = {iN^2 \over
m} \pmatrix{0&-p_2& p_1& 0\cr p_2 &0& m+{p_r p_l\over p_0+m} & {p_2
p_3\over p_0 +m}\cr -p_1 & -m - {p_r p_l \over p_0+m}& 0& -{p_1 p_3\over
p_0 +m}\cr 0& -{p_2 p_3 \over p_0 +m} & {p_1 p_3 \over p_0+m}&0\cr}\, ,
\label{lc} \end{eqnarray}
which coincides with the `longitudinal' components
of the  antisymmetric tensor which have been obtained in previous 
works of ours within normalization and different forms of the
spinorial basis.  The longitudinal states can be eliminated in
the massless case when one uses suitable
normalization and if a $S=1$ particle moves
along the third-axis $OZ$ direction. It is also useful to compare Eq.
(\ref{lc}) with the formula (B2) in ref.~\cite{Ahl-lf}, the expressions
for the strengths in the light-front form of the QFT, in
order to realize the correct procedure for taking the massless limit.

As a discussion we want to mention that the Tam-Happer experiment~\cite{TH}
did not find a satisfactory explanation in the
quantumelectrodynamic frameworks
(at least, its explanation is complicated
by tedious technical calculus). On the other hand, in
ref.~\cite{Prad} an interesting
model has been proposed.  It is based on gauging the Dirac field
on using a set of parameters which are dependent on
space-time coordinates $\alpha_{\mu\nu} (x) $ in $\psi (x)\rightarrow
\psi^\prime (x^\prime)=\Omega\psi (x),\,\,\quad\Omega=\exp\left [
{i\over 2}\sigma^{\mu\nu}\alpha_{\mu\nu} (x)\right ]\, $.
Thus, the second ``photon" has been taken into consideration.
The 24-component compensation field  $B_{\mu,\nu\lambda} $
reduces to the 4-vector field as follows
(the notation of~\cite{Prad} is used here):  $B_{\mu,\nu\lambda}=
{1\over 4}\epsilon_{\mu\nu\lambda\sigma} a_\sigma (x)\, .$
As you can readily see after the
comparison of the formulas of~\cite{Prad} with those of
refs.~\cite{Og,Og1,Og2}, the second photon of Pradhan and Naik is nothing
more than the Ogievetski\u{\i}-Polubarinov {\it notoph} within the
normalization.  Parity properties (massless behavior as well) are the
matter of dependence, not only on the explicit forms  of the field
functions in the momentum space $ (1/2,1/2)$ representation, {\it but}
also on the properties of the corresponding creation/annihilation
operators.  The helicity properties  in the massless limit depend on the
normalization.

Moreover, it appears that the properties of the polarization
vectors with respect to parity operation depend on the choice of the spin basis.
For instance, in Ref.~\cite{dv-cl,GR} the momentum-space polarization vectors have been 
listed in the helicity basis:
\begin{eqnarray}
&&\epsilon _{\mu }({\bf p},\lambda =+1)={\frac{1}{\sqrt{2}}}{\frac{e^{i\phi }}{
p}}
\pmatrix{0, {p_x p_z -ip_y p\over \sqrt{p_x^2 +p_y^2}}, {p_y p_z +ip_x
p\over \sqrt{p_x^2 +p_y^2}}, -\sqrt{p_x^2 +p_y^2}},\\
&&\epsilon _{\mu }({\bf p},\lambda =-1)={\frac{1}{\sqrt{2}}}{\frac{e^{-i\phi }
}{p}}\pmatrix{ 0, {-p_x p_z -ip_y p\over \sqrt{p_x^2 +p_y^2}}, {-p_y p_z
+ip_x p\over \sqrt{p_x^2 +p_y^2}}, +\sqrt{p_x^2 +p_y^2}},\nonumber\\
\\
&&\epsilon _{\mu }({\bf p},\lambda =0)={\frac{1}{m}}\pmatrix{ p, -{E \over p}
p_x, -{E \over p} p_y, -{E \over p} p_z }\,, \\
&&\epsilon _{\mu }({\bf p},\lambda =0_{t})={\frac{1}{m}}\pmatrix{E , -p_x,
-p_y, -p_z }\,.
\end{eqnarray}
Berestetski\u{\i}, Lifshitz and Pitaevski\u{\i} claimed too, Ref.~\cite{BLP}, that
the helicity states cannot be the parity states. If one applies the common-used
relations between fields and potentials it appears that the ${\bf E}$ and ${\bf B}$ fields have no usual properties with respect to space inversions:
\begin{eqnarray}
&&{\bf E}({\bf p},\lambda =+1)=-{\frac{iEp_{z}}{\sqrt{2}pp_{l}}}{\bf p}-{\frac{
E}{\sqrt{2}p_{l}}}\tilde{{\bf p}},\, {\bf B}({\bf p},\lambda =+1)={\frac{
p_{z}}{\sqrt{2}p_{l}}}{\bf p}-{\frac{ip}{\sqrt{2}p_{l}}}\tilde{{\bf p}}, \nonumber\\
&&\\
&&{\bf E}({\bf p},\lambda =-1)=+{\frac{iEp_{z}}{\sqrt{2}pp_{r}}}{\bf p}-{\frac{
E}{\sqrt{2}p_{r}}}\tilde{{\bf p}}^{\ast },\, {\bf B}({\bf p},\lambda
=-1)={\frac{p_{z}}{\sqrt{2}p_{r}}}{\bf p}+{\frac{ip}{\sqrt{2}p_{r}}}\tilde{
{\bf p}}^{\ast }, \nonumber\\
&&\\
&&{\bf E}({\bf p},\lambda =0)={\frac{im}{p}}{\bf p},\quad {\bf B}({\bf p}
,\lambda =0)=0,
\end{eqnarray}
with $\tilde {\bf p}=\pmatrix{p_y\cr -p_x\cr -ip\cr}$.

Thus, the conclusions of the previous works are:
\begin{itemize}

\item
The mapping exists between the WTH formalism for $S=1$ and the AST fields of four kinds (provided that the solutions of the WTH equations are of the definite
parity).

\item
Their massless limits contain additional solutions comparing with the Maxwell equations. This was related to the possible theoretical existence of the Ogievetski\u{\i}-Polubarinov-Kalb-Ramond notoph, Ref.~\cite{Og,Og1,Og2}.

\item
In some particular cases ($A=0, B=1$) massive solutions of different parities are naturally divided into the classes of causal and tachyonic solutions.

\item
If we want to take into account the solutions of the WTH equations of
different parity properties, this induces us to generalize the BW, Proca and Duffin-Kemmer formalisms.

\item
In the $(1/2,0)\oplus (0,1/2)$, $(1,0)\oplus (0,1)$ etc. representations
it is possible to introduce the parity-violating frameworks. The corresponding solutions are the mixing of various polarization states.

\item
The sum of the Klein-Gordon equation with the $(S,0)\oplus (0,S)$ equations may change the theoretical content even on the free level. For instance, the higher-spin equations may actually describe various spin and mass states.

\item
The mappings exists between the WTH solutions of undefined parity
and the AST fields, which contain both tensor and dual tensor. They are eight.

\item
The 4-potentials and electromagnetic fields~\cite{dv-cl,GR} in the helicity
basis have different parity properties comparing with the standard basis of the polarization vectors.

\item
In the previous talk~\cite{dv-pl} I presented a theory in the $(1/2,0)\oplus (0,1/2)$ representation in the helicity basis. Under the space inversion operation,
different helicity states transform each other, $Pu_h (-{\bf p}) = -i u_{-h} ({\bf p})$, $Pv_h (-{\bf p}) = +i v_{-h} ({\bf p})$.

\end{itemize}

\section{The theory of 4-vector field.}

First of all, we show that the equation for the 4-vector field can be presented
in a matrix form. Recently, S. I. Kruglov proposed, Ref.~\cite{krug1}, 
a general form of the Lagrangian for  4-potential field $B_\mu$, which also contains the spin-0 state. 
Initially, we have 
\begin{equation}
\alpha \partial_\mu \partial_\nu B_\nu +\beta \partial_\nu^2 B_\mu +\gamma m^2 B_\mu =0\, ,\label{eq-pot}
\end{equation}
provided that derivatives commute.
When $\partial_\nu B_\nu =0$ (the Lorentz gauge) we obtain the spin-1 states only.
However, if it is not equal to zero we have a scalar field and an axial-vector potential. We can also verify this 
statement by consideration of the dispersion relations of the equation (\ref{eq-pot}). One obtains 4+4 states 
(two of them may differ in mass from others).

Next, one can fix one of the constants $\alpha,\beta,\gamma$ 
without loosing any physical content. For instance, when $\alpha=-2$
one gets the equation
\begin{equation}
\left [ \delta_{\mu\nu} \delta_{\alpha\beta} - \delta_{\mu\alpha}\delta_{\nu\beta} - \delta_{\mu\beta} \delta_{\nu\alpha}\right ] \partial_\alpha \partial_\beta B_\nu + A \partial_\alpha^2 \delta_{\mu\nu} B_\nu - Bm^2  B_\mu =0\,,\label{eq1-m}
\end{equation} 
where  $\beta= A+1$ and $\gamma=-B$. In the matrix form the equation (\ref{eq1-m}) reads:
\begin{equation}
\left [ \gamma_{\alpha\beta} \partial_\alpha \partial_\beta +A \partial_\alpha^2 - Bm^2 \right ]_{\mu\nu} B_\nu = 0\,,
\end{equation}
with
\begin{equation}
[\gamma_{\alpha\beta}]_{\mu\nu} = \delta_{\mu\nu}\delta_{\alpha\beta}
-\delta_{\mu\alpha}\delta_{\nu\beta} - \delta_{\mu\beta} \delta_{\nu\alpha}\,.
\end{equation}
Their explicit forms are the following ones:
\begin{eqnarray}
&&\gamma_{44}=\pmatrix{1&0&0&0\cr
0&1&0&0\cr
0&0&1&0\cr
0&0&0&-1\cr}\,,\quad
\gamma_{14}=\gamma_{41}=\pmatrix{0&0&0&-1\cr
0&0&0&0\cr
0&0&0&0\cr
-1&0&0&0\cr}\,,\\
&&\gamma_{24}=\gamma_{42}=\pmatrix{0&0&0&0\cr
0&0&0&-1\cr
0&0&0&0\cr
0&-1&0&0\cr}\,,\quad
\gamma_{34}=\gamma_{43}=\pmatrix{0&0&0&0\cr
0&0&0&0\cr
0&0&0&-1\cr
0&0&-1&0\cr}\nonumber\\
\\
&&\gamma_{11}=\pmatrix{-1&0&0&0\cr
0&1&0&0\cr
0&0&1&0\cr
0&0&0&1\cr}\,,\quad
\gamma_{22}=\pmatrix{1&0&0&0\cr
0&-1&0&0\cr
0&0&1&0\cr
0&0&0&1\cr}\,,\\
&&\gamma_{33}=\pmatrix{1&0&0&0\cr
0&1&0&0\cr
0&0&-1&0\cr
0&0&0&1\cr}\,,\quad
\gamma_{12}=\gamma_{21}=\pmatrix{0&-1&0&0\cr
-1&0&0&0\cr
0&0&0&0\cr
0&0&0&0\cr}\,,\\
&&\gamma_{31}=\gamma_{13}=\pmatrix{0&0&-1&0\cr
0&0&0&0\cr
-1&0&0&0\cr
0&0&0&0\cr}\,,\quad
\gamma_{23}=\gamma_{32}=\pmatrix{0&0&0&0\cr
0&0&-1&0\cr
0&-1&0&0\cr
0&0&0&0\cr}.\nonumber\\
\end{eqnarray}
They are the analogs of the Barut-Muzinich-Williams (BMW) $\gamma$-matrices
for bivector fields. However, $\sum_{\alpha}^{} [\gamma_{\alpha\alpha}]_{\mu\nu} = 
2\delta_{\mu\nu}$. One can also define the analogs of the BMW $\gamma_{5,\alpha\beta}$ matrices 
\begin{equation}
\gamma_{5,\alpha\beta} = \frac{i}{6} [\gamma_{\alpha\kappa}, \gamma_{\beta\kappa} ]_{-, \mu\nu} = i [\delta_{\alpha\mu} \delta_{\beta\nu} - \delta_{\alpha\nu}\delta_{\beta\mu} ]\,.
\end{equation}
As opposed to $\gamma_{\alpha\beta}$ matrices they are totally antisymmetric.
They are related to  boost and rotation generators of this representation.
Their explicite forms are:
\begin{eqnarray}
&&\gamma_{5,41} = - \gamma_{5,14} = i \pmatrix{0&0&0&-1\cr
0&0&0&0\cr
0&0&0&0\cr
1&0&0&0\cr}\,,\,
\gamma_{5,42} = - \gamma_{5,24} = i \pmatrix{0&0&0&0\cr
0&0&0&-1\cr
0&0&0&0\cr
0&1&0&0\cr}\,,\nonumber\\
&&\\
&&\nonumber\\
&&\gamma_{5,43} = - \gamma_{5,34} = i \pmatrix{0&0&0&0\cr
0&0&0&0\cr
0&0&0&-1\cr
0&0&1&0\cr}\,,\,
\gamma_{5,12} = - \gamma_{5,21} = i \pmatrix{0&1&0&0\cr
-1&0&0&0\cr
0&0&0&0\cr
0&0&0&0\cr}\,,\nonumber\\
&&\\
&&\gamma_{5,31} = - \gamma_{5,13} = i \pmatrix{0&0&-1&0\cr
0&0&0&0\cr
1&0&0&0\cr
0&0&0&0\cr}\,,\,
\gamma_{5,23} = - \gamma_{5,32} = i \pmatrix{0&0&0&0\cr
0&0&1&0\cr
0&-1&0&0\cr
0&0&0&0\cr}\,.\nonumber\\
&&
\end{eqnarray}
The $\gamma$-matrices are pure real; the $\gamma_5$-matrices are pure imaginary.
In the $(1/2,1/2)$ representation, we need 16 matrices to form the complete set.
It is easy to prove by the textbook method~\cite{Itzyk} that $\gamma_{44}$
can serve as the parity matrix.

{\it Lagrangian and the equations of motion.}
Let us try
\begin{equation}
{\cal L} = (\partial_\alpha B_\mu^\ast) [\gamma_{\alpha\beta} ]_{\mu\nu} (\partial_\beta B_\nu)
+ A (\partial_\alpha B_\mu^\ast) (\partial_\alpha B_\mu) + Bm^2
B_\mu^\ast B_\mu\,.
\end{equation}
On using the Lagrange-Euler equation
\begin{equation}
\frac{\partial {\cal L}}{\partial B_\mu^\ast}
- \partial_\nu (\frac{\partial {\cal L}}{\partial (\partial_\nu B_\mu^\ast)})
=0\,,
\end{equation}
or
\begin{equation}
\frac{\partial {\cal L}}{\partial B_\mu}
- \partial_\nu (\frac{\partial {\cal L}}{\partial (\partial_\nu B_\mu)})
=0\,,
\end{equation}
we have
\begin{equation}
[\gamma_{\nu\beta}]_{\kappa\tau} \partial_\nu \partial_\beta B_\tau + A \partial_\nu^2 B_\kappa - Bm^2  B_\kappa =0\,,\label{equat}
\end{equation}
or
\begin{equation}
[\gamma_{\beta\nu}]_{\kappa\tau} \partial_\beta \partial_\nu B_\tau^\ast + A \partial_\nu^2 B_\kappa^\ast - Bm^2  B_\kappa^\ast =0\,.\label{eq-f}
\end{equation}

\medskip

{\it Masses.} We are convinced that in the case of spin 0, we have $B_\mu \rightarrow
\partial_\mu \chi$; in the case of spin 1 we have $\partial_\mu B_\mu =0$.

So,
\begin{enumerate}

\item

\begin{equation}
(\delta_{\mu\nu}\delta_{\alpha\beta}
-\delta_{\mu\alpha}\delta_{\nu\beta} - \delta_{\mu\beta} \delta_{\nu\alpha})
\partial_\alpha \partial_\beta \partial_\nu \chi = - \partial^2 \partial_\mu \chi\,.
\end{equation}
Hence, from (\ref{equat}) we have
\begin{equation}
[(A-1) \partial^2_\nu - Bm^2 ] \partial_\mu \chi=0\,.
\end{equation}
If $A-1=B$ we have the spin-0 particles with masses $\pm m$ with the correct relativistic dispersion.

\smallskip

\item
In another case
\begin{equation}
[\delta_{\mu\nu}\delta_{\alpha\beta}
-\delta_{\mu\alpha}\delta_{\nu\beta} - \delta_{\mu\beta} \delta_{\nu\alpha}]
\partial_\alpha \partial_\beta B_\nu  =  \partial^2 B_\mu \,.
\end{equation}
Hence,
\begin{equation}
[(A+1) \partial^2_\nu  - Bm^2] B_\mu =0\,.
\end{equation}
If $A+1 =B$ we have the spin-1 particles with masses $\pm m$ with the correct relativistic dispersion.

\end{enumerate}

The equation (\ref{equat}) can be transformed in  two equations:
\begin{eqnarray}
\left [\gamma_{\alpha\beta} \partial_\alpha \partial_\beta + (B+1)\partial_\alpha^2 - Bm^2 \right ]_{\mu\nu} B_\nu &=&0\,, \quad \mbox{spin 0 with masses}\, \pm m\nonumber\\
\\
\left [\gamma_{\alpha\beta} \partial_\alpha \partial_\beta + (B-1)\partial_\alpha^2  - Bm^2 \right ]_{\mu\nu}  B_\nu &=&0\,, \quad \mbox{spin 1 with masses}\, \pm m.\nonumber\\
\end{eqnarray}
The first one has the solution with spin 0 and masses $\pm m$. However, it has also the {\it spin-1} solution with the {\it different masses}, $[\partial_\nu^2 +(B+1)\partial^2_\nu - Bm^2 ] B_\mu =0$:
\begin{equation}
\tilde m = \pm \sqrt{{B\over B+2}} m\,.
\end{equation}
The second one has the solution with spin 1 and masses $\pm m$. But, it also has 
the {\it spin-0} solution with the {\it different masses}, $[ -\partial_\nu^2 + (B-1) \partial^2_\nu - Bm^2 ] \partial_\mu \chi =0$:
\begin{equation}
\tilde m = \pm \sqrt{{B\over B-2}}m\,.
\end{equation}
One can come to the same conclusion by checking the dispersion relations 
from $\mbox{Det} [\gamma_{\alpha\beta} p_\alpha p_\beta - Ap_\alpha p_\alpha +Bm^2] = 0$\,. When $\tilde m^2 = {4\over 3} m^2$, we have $B=-8, A=-7$, that is compatible with our consideration of bi-vector fields~\cite{wig}.

One can form the Lagrangian with the particles of spines 1, masses $\pm m$, the particle with the mass $\sqrt{{4\over 3}} m$, spin 1, for which the particle is equal to the antiparticle, by choosing the appropriate creation/annihilation operators; and the particles with spines 0 with masses $\pm m$ and 
$\pm \sqrt{{4\over 5}} m$ (some of them may be neutral).

\medskip

{\it The Standard Basis}~\cite{Novozh,Weinb,Dv-book}.
The polarization vectors of the standard basis are defined:
\begin{eqnarray}
&&\epsilon_\mu
({\bf 0}, +1)= -{1\over \sqrt{2}}\pmatrix{1\cr i\cr 0 \cr 0\cr}\,,\, 
\epsilon_\mu ({\bf 0}, -1)= +{1\over
\sqrt{2}}\pmatrix{1\cr -i\cr 0\cr 0\cr},\\
&&\epsilon_\mu ({\bf
0}, 0) = \pmatrix{0\cr 0\cr 1\cr 0\cr}\,,\,
\epsilon_\mu ({\bf
0}, 0_t) = \pmatrix{0\cr 0\cr 0\cr i\cr}.
\end{eqnarray}
The Lorentz transformations are:
\begin{eqnarray} &&
\epsilon_\mu ({\bf p}, \sigma) =
L_{\mu\nu} ({\bf p}) \epsilon_\nu ({\bf 0},\sigma)\,,\\ 
&& L_{44} ({\bf p}) = \gamma\, ,\, L_{i4} ({\bf p}) = -
L_{4i} ({\bf p}) = i\widehat p_i \sqrt{\gamma^2 -1}\, ,\,
L_{ik} ({\bf p}) = \delta_{ik} + (\gamma -1) \widehat p_i \widehat
p_k .\nonumber\\ 
\end{eqnarray} 
Hence, for the particles of the mass $m$ we have:
\begin{eqnarray} 
u^\mu
({\bf p}, +1)&=& -{N\over \sqrt{2}m}\pmatrix{m+ {p_1 p_r \over
E_p+m}\cr im +{p_2 p_r \over E_p+m}\cr {p_3 p_r \over
E_p+m}\cr -ip_r\cr}\,,\\
 u^\mu ({\bf p}, -1)&=& {N\over
\sqrt{2}m}\pmatrix{m+ {p_1 p_l \over E_p+m}\cr -im +{p_2 p_l \over
E_p+m}\cr {p_3 p_l \over E_p+m}\cr -ip_l\cr}\,,\label{vp12}\\ 
u^\mu ({\bf
p}, 0) &=& {N\over m}\pmatrix{{p_1 p_3 \over E_p+m}\cr {p_2 p_3
\over E_p+m}\cr m + {p_3^2 \over E_p+m}\cr -ip_3\cr}\,,\quad
u^\mu ({\bf p}, 0_t) = {N \over m} \pmatrix{-p_1
\cr -p_2\cr -p_3\cr iE_p\cr}\,.
\end{eqnarray}
The Euclidean metric was again used; $N$ is the normalization constant. They are the eigenvectors of the parity operator:
\begin{equation}
Pu_\mu (-{\bf p}, \sigma) = + u_\mu ({\bf p}, \sigma)\,,\quad
Pu_\mu (-{\bf p}, 0_t) = -u_\mu ({\bf p}, 0_t)\,.
\end{equation}

\medskip

{\it The Helicity Basis.}~\cite{GR,Car}
The helicity operator is:
\begin{eqnarray}
&&{({\bf J}\cdot {\bf p})\over p} = {1\over p} \pmatrix{
0&-ip_z&ip_y&0\cr
ip_z&0&-ip_x&0\cr
-ip_y&ip_x&0&0\cr
0&0&0&0\cr}\,,\,\\
&&{({\bf J}\cdot {\bf p})\over p} \epsilon^\mu_{\pm 1} = \pm 
\epsilon^\mu_{\pm 1}\,,\,\,{({\bf J}\cdot {\bf p})\over p} \epsilon^\mu_{0,0_t} = 0\,.
\end{eqnarray}
The eigenvectors are:
\begin{eqnarray}
&&\epsilon^\mu_{+1}= {1\over \sqrt{2}} {e^{i\alpha}\over p} \pmatrix{
{-p_x p_z +ip_y p\over \sqrt{p_x^2 +p_y^2}}\cr {-p_y p_z -ip_x
p\over \sqrt{p_x^2 +p_y^2}}\cr \sqrt{p_x^2 +p_y^2}\cr 0\cr}\,,\,
\epsilon^{\mu }_{-1}={1\over \sqrt{2}}{e^{i\beta}\over p}
\pmatrix{{p_x p_z +ip_y p\over \sqrt{p_x^2 +p_y^2}}\cr {p_y p_z
-ip_x p\over \sqrt{p_x^2 +p_y^2}}\cr -\sqrt{p_x^2 +p_y^2}\cr 0\cr},\nonumber\\ 
\\
&&\epsilon^{\mu }_0={\frac{1}{m}}\pmatrix{{E \over p}
p_x \cr {E \over p} p_y \cr{E \over p} p_z\cr ip }\,,\quad
\epsilon^{\mu }_{0_{t}}={\frac{1}{m}}\pmatrix{p_x\cr
p_y\cr p_z\cr iE_p\cr}\,.
\end{eqnarray}
The eigenvectors $\epsilon^\mu_{\pm  1}$ are not the eigenvectors
of the parity operator ($\gamma_{44}$) of this representation. However, $\epsilon^\mu_{1,0}$,
$\epsilon^\mu_{0,0_t}$
are. Surprisingly, the latter have no well-defined massless limit.\footnote{In order to get the well-known massless limit one should use the basis of the light-front form reprersentation, ref~\cite{Ahl-lf}.}

\medskip

{\it Energy-momentum tensor.}
According to definitions~\cite{Lurie} it is defined as
\begin{eqnarray}
&&T_{\mu\nu} = - \sum_{\alpha}^{} \left [ {\partial {\cal L} \over \partial (\partial_\mu B_\alpha)} \partial_\nu B_\alpha
+\partial_\nu B_\alpha^\ast {\partial {\cal L} \over \partial (\partial_\mu B_\alpha^\ast)}\right ]
+{\cal L} \delta_{\mu\nu},\nonumber\\
\\
&& P_\mu = -i \int T_{4\mu} d^3 {\bf x}\,.
\end{eqnarray}
Hence,
\begin{eqnarray}
&&T_{\mu\nu} =  -(\partial_\kappa B_\tau^\ast) [\gamma_{\kappa\mu}]_{\tau\alpha} (\partial_\nu B_\alpha) - (\partial_\nu B_\alpha^\ast) [\gamma_{\mu\kappa}]_{\alpha\tau} (\partial_\kappa B_\tau)-\nonumber\\
&-& A [(\partial_\mu B_\alpha^\ast) (\partial_\nu B_\alpha) + (\partial_\nu B_\alpha^\ast)  (\partial_\mu B_\alpha)] + {\cal L}\delta_{\mu\nu} =\nonumber\\
&=& - (A+1) [(\partial_\mu B_\alpha^\ast) (\partial_\nu B_\alpha) + (\partial_\nu B_\alpha^\ast)  (\partial_\mu B_\alpha)] +
\left [ (\partial_\alpha B_\mu^\ast) (\partial_\nu B_\alpha) + \right . \nonumber\\
&+&\left . (\partial_\nu B_\alpha^\ast)  (\partial_\alpha B_\mu) \right ] +
[(\partial_\alpha B_\alpha^\ast) (\partial_\nu B_\mu) + (\partial_\nu B_\mu^\ast)  (\partial_\alpha B_\alpha)] + {\cal L} \delta_{\mu\nu}.
\end{eqnarray}
Remember that after substitutions of  the explicite forms $\gamma$'s, the Lagrangian is
\begin{equation}
{\cal L} = (A+1) (\partial_\alpha B_\mu^\ast) (\partial_\alpha B_\mu ) - (\partial_\nu B_\mu^\ast)  (\partial_\mu B_\nu)- (\partial_\mu B_\mu^\ast) (\partial_\nu B_\nu) + Bm^2   B_\mu^\ast B_\mu\,,
\end{equation}
and the third term cannot be removed by the standard substitution ${\cal L} \rightarrow {\cal L}^\prime +\partial_\mu \Gamma_\mu$\,,$\Gamma_\mu = B_\nu^\ast \partial_\nu B_\mu - B_\mu^\ast \partial_\nu B_\nu$ 
to get the textbook Lagrangian ${\cal L}^\prime = (\partial_\alpha B_\mu^\ast) (\partial_\alpha B_\mu ) +m^2 B_\mu^\ast B_\mu$\,.

\medskip

{\it The current vector} is defined
\begin{eqnarray}
&&J_{\mu} = -i \sum_{\alpha}^{} [{\partial {\cal L} \over \partial (\partial_\mu B_\alpha)} 
B_\alpha
- B_\alpha^\ast {\partial {\cal L} \over \partial (\partial_\mu B_\alpha^\ast)} ]\,,\\
&& Q = -i \int J_{4} d^3 {\bf x}\,.
\end{eqnarray}
Hence,
\begin{eqnarray}
&&J_{\lambda} =  -i \left \{ (\partial_\alpha B_\mu^\ast) [\gamma_{\alpha\lambda}]_{\mu\kappa}  B_\kappa - B_\kappa^\ast [\gamma_{\lambda\alpha}]_{\kappa\mu} (\partial_\alpha B_\mu) +\right.\nonumber\\
&&\left . +A (\partial_\lambda B_\kappa^\ast) B_\kappa - A B_\kappa^\ast (\partial_\lambda B_\kappa) \right \} = \nonumber\\
&=& - i \left \{ (A+1) [(\partial_\lambda B_\kappa^\ast) B_\kappa
 -  B_\kappa^\ast (\partial_\lambda B_\kappa) ] +  [ B_\kappa^\ast (\partial_\kappa B_\lambda) -
(\partial_\kappa B_\lambda^\ast) B_\kappa ] + \right . \nonumber\\
&+&  \left . [B_\lambda^\ast (\partial_\kappa B_\kappa) - (\partial_\kappa B_\kappa^\ast) B_\lambda ] \right \} \,.
\end{eqnarray}
Again, the second term and the last term cannot be removed at the same time by adding the total derivative to the Lagrangian. These terms correspond to the contribution of the scalar (spin-0) portion.

\medskip

{\it Angular momentum.}
Finally,
\begin{eqnarray}
&&{\cal M}_{\mu\alpha,\lambda} = x_\mu T_{\{\alpha\lambda\}} - x_\alpha T_{\{\mu\lambda\}} + {\cal S}_{\mu\alpha,\lambda} =\nonumber\\
&=& x_\mu T_{\{\alpha\lambda\}} - x_\alpha T_{\{\mu\lambda\}} -  i \left \{\sum_{\kappa\tau}^{}
{\partial {\cal L} \over \partial (\partial_\lambda B_\kappa)} {\cal T}_{\mu\alpha,\kappa\tau} B_\tau+ B_\tau^\ast {\cal T}_{\mu\alpha,\kappa\tau} {\partial {\cal L} \over 
\partial (\partial_\lambda B_\kappa^\ast)}\right \},\nonumber\\
\\
&& {\cal M}_{\mu\nu} = -i \int {\cal M}_{\mu\nu,4} d^3 {\bf x}\,,
\end{eqnarray}
where ${\cal T}_{\mu\alpha,\kappa\tau} \sim [\gamma_{5,\mu\alpha}]_{\kappa\tau}$\,.

\medskip

{\it The field operator.} Various-type field operators are possible in this representation. Let us remind the textbook procedure to get them.
During the calculations below we have to present $1=\theta (k_0) +\theta (-k_0)$
in order to get positive- and negative-frequency parts. However, one should be warned that in the point $k_0=0$ this presentation is ill-defined.
\begin{eqnarray}
&&A_\mu (x) = {1\over (2\pi)^3} \int d^4 k \,\delta (k^2 -m^2) e^{+ik\cdot x}
A_\mu (k) =\nonumber\\
&=& {1\over (2\pi)^3} \sum_{\lambda}^{}\int d^4 k \delta (k_0^2 -E_k^2) e^{+ik\cdot x}
\epsilon_\mu (k,\lambda) a_\lambda (k) =\nonumber\\
&=&{1\over (2\pi)^3} \int {d^4 k \over 2E_k} [\delta (k_0 -E_k) +\delta (k_0 +E_k) ] 
[\theta (k_0) +\theta (-k_0) ]e^{+ik\cdot x}
A_\mu (k) =\nonumber\\
&=&{1\over (2\pi)^3} \int {d^4 k \over 2E_k} [\delta (k_0 -E_k) +\delta (k_0 +E_k) ] \left
[\theta (k_0) A_\mu (k) e^{+ik\cdot x}  + \right.\nonumber\\
&+&\left.\theta (k_0) A_\mu (-k) e^{-ik\cdot x} \right ]
={1\over (2\pi)^3} \int {d^3 {\bf k} \over 2E_k} \theta(k_0)  
[A_\mu (k) e^{+ik\cdot x}  + A_\mu (-k) e^{-ik\cdot x} ]
=\nonumber\\
&=&{1\over (2\pi)^3} \sum_{\lambda}^{}\int {d^3 {\bf k} \over 2E_k}   
[\epsilon_\mu (k,\lambda) a_\lambda (k) e^{+ik\cdot x}  + \epsilon_\mu (-k,\lambda) 
a_\lambda (-k) e^{-ik\cdot x} ]\,.
\end{eqnarray}
Moreover, we should transform the second part to $\epsilon_\mu^\ast (k,\lambda) b_\lambda^\dagger (k)$ as usual. In such a way we obtain the charge-conjugate states. Of course, one can try to get $P$-conjugates or $CP$-conjugate states too. 
One should proceed as in the spin-1/2 case.

In the Dirac case we should assume the following relation in the field operator:
\begin{equation}
\sum_{\lambda}^{} v_\lambda (k) b_\lambda^\dagger (k) = \sum_{\lambda}^{} u_\lambda (-k) a_\lambda (-k)\,.\label{dcop}
\end{equation}
We know that~\cite{Ryder}
\begin{eqnarray}
\bar u_\mu (k) u_\lambda (k) &=& +m \delta_{\mu\lambda}\,,\\
\bar u_\mu (k) u_\lambda (-k) &=& 0\,,\\
\bar v_\mu (k) v_\lambda (k) &=& -m \delta_{\mu\lambda}\,,\\
\bar v_\mu (k) u_\lambda (k) &=& 0\,,
\end{eqnarray}
but we need $\Lambda_{\mu\lambda} (k) = \bar v_\mu (k) u_\lambda (-k)$.
By direct calculations,  we find
\begin{equation}
-mb_\mu^\dagger (k) = \sum_{\nu}^{} \Lambda_{\mu\lambda} (k) a_\lambda (-k)\,.
\end{equation}
Hence, $\Lambda_{\mu\lambda} = -im ({\bf \sigma}\cdot {\bf n})_{\mu\lambda}$
and 
\begin{equation}
b_\mu^\dagger (k) = i({\bf \sigma}\cdot {\bf n})_{\mu\lambda} a_\lambda (-k)\,.
\end{equation}
Multiplying (\ref{dcop}) by $\bar u_\mu (-k)$ we obtain
\begin{equation}
a_\mu (-k) = -i ({\bf \sigma} \cdot {\bf n})_{\mu\lambda} b_\lambda^\dagger (k)\,.
\end{equation}
The equations (60) and (61) are self-consistent.

In the $(1,0)\oplus (0,1)$ representation we have somewhat different situation:
\begin{equation}
a_\mu (k) = [1-2({\bf S}\cdot {\bf n})^2]_{\mu\lambda} a_\lambda (-k)\,. 
\end{equation}
This signifies that in order to construct the Sankaranarayanan-Good field operator, which was used by Ahluwalia, Johnson and Goldman, it satisfies 
$[\gamma_{\mu\nu} \partial_\mu \partial_\nu - {(i\partial/\partial t)\over E} 
m^2 ] \Psi =0$, we need additional postulates.

We set in the $(1/2,1/2)$ case:
\begin{equation}
\sum_{\lambda}^{} \epsilon_\mu (-k,\lambda) a_\lambda (-k) = 
\sum_{\lambda}^{} \epsilon_\mu^\ast (k,\lambda) b_\lambda^\dagger (k)\,,
\label{expan}
\end{equation}
multiply both parts by $\epsilon_\nu [\gamma_{44}]_{\nu\mu}$, and use the normalization conditions for polarization vectors.

In the $({1\over 2}, {1\over 2})$ representation we can also expand
(apart the equation (\ref{expan})) in the different way:
\begin{equation}
\sum_{\lambda}^{} \epsilon_\mu (-k, \lambda) a_\lambda (-k) =
\sum_{\lambda}^{} \epsilon_\mu (k, \lambda) a_\lambda (k)\,.
\end{equation}
From the first definition we obtain (the signs $\mp$
depends on the value of $\sigma$):
\begin{equation}
b_\sigma^\dagger (k) = \mp \sum_{\mu\nu\lambda}^{} \epsilon_\nu (k,\sigma) 
[\gamma_{44}]_{\nu\mu} \epsilon_\mu (-k,\lambda) a_\lambda (-k)\,,
\end{equation}
or
\begin{eqnarray}
b_\sigma^\dagger (k) = {E_k^2 \over m^2} \pmatrix{1+{{\bf k}^2\over E_k^2}&\sqrt{2}
{k_r \over E_k}&-\sqrt{2} {k_l \over E_k}& -{2k_3 \over E_k}\cr
-\sqrt{2} {k_r \over E_k}&-{k_r^2 \over {\bf k}^2}& -{m^2k_3^2\over E_k^2 {\bf k}^2}
+{k_r k_l \over E_k^2} & {\sqrt{2} k_3 k_r \over {\bf k}^2}\cr
\sqrt{2} {k_l \over E_k}&-{m^2 k_3^2 \over E_k^2 {\bf k}^2} + {k_r k_l \over E_k^2}& -{k_l^2\over {\bf k}^2} & -{\sqrt{2} k_3 k_l \over {\bf k}^2}\cr
{2k_3 \over E_k}&{\sqrt{2}k_3 k_r \over {\bf k}^2}& -{\sqrt{2} k_3 k_l\over {\bf k}^2} & {m^2 \over E_k^2} -{2 k_3 \over {\bf k}^2}\cr}
\pmatrix{a_{00} (-k)\cr a_{11} (-k)\cr
a_{1-1} (-k)\cr a_{10} (-k)\cr}\,.\nonumber
\end{eqnarray}
\begin{equation}
.
\end{equation}
From the second definition $\Lambda^2_{\sigma\lambda} = \mp \sum_{\nu\mu}^{} \epsilon^{\ast}_\nu (k, \sigma) [\gamma_{44}]_{\nu\mu}
\epsilon_\mu (-k, \lambda)$ we have
\begin{eqnarray}
a_\sigma (k) =  \pmatrix{-1&0&0&0\cr
0&{k_3^2 \over {\bf k}^2}& {k_l^2\over {\bf k}^2} & {\sqrt{2} k_3 k_l \over {\bf k}^2}\cr
0&{k_r^2 \over {\bf k}^2}& {k_3^2\over {\bf k}^2} & -{\sqrt{2} k_3 k_r \over {\bf k}^2}\cr
0&{\sqrt{2}k_3 k_r \over {\bf k}^2}& -{\sqrt{2} k_3 k_l\over {\bf k}^2} & 1-{2 k_3^2 \over {\bf k}^2}\cr}\pmatrix{a_{00} (-k)\cr a_{11} (-k)\cr
a_{1-1} (-k)\cr a_{10} (-k)\cr}\,.
\end{eqnarray}
It is the strange case: the field operator will only destroy particles. Possibly, we should think about modifications of the Fock space in this case, or introduce several field operators for the $({1\over 2}, {1\over 2})$ representation.

\medskip

{\it Propagators.} From ref.~\cite{Itzyk} it is known for the real vector field:
\begin{equation}
<0\vert T(B_\mu (x) B_\nu (y)\vert 0> = -i \int {d^4 k \over (2\pi)^4} e^{ik (x-y)} 
(\frac{\delta_{\mu\nu} +k_\mu k_\nu/\mu^2}{k^2 +\mu^2 +i\epsilon} - \frac{k_\mu k_\nu/\mu^2}{k^2 +m^2 +i\epsilon})\,.
\end{equation}
If $\mu=m$ (this depends on relations between $A$ and  $B$) we have the cancellation of divergent parts. Thus, we can overcome the well-known  difficulty of the Proca theory with the massless limit. 

If $\mu\neq m$ we can still have a {\it causal} theory, but in this case we need more than one equation, and should apply the method proposed 
in ref.~\cite{dv-hpa}.\footnote{In that case we applied  for the bi-vector fields
\begin{eqnarray}
&&\hspace*{-1cm}\left [ \gamma_{\mu\nu} \partial_\mu \partial_\nu -m^2 \right ]
 \int  \frac{d^3 {\bf p}}{(2\pi)^3 8im^2 E_p}
\left [ \theta (t_2 -t_1) u^1_{\sigma} ({\bf p}) \otimes \overline
u^1_{\sigma} ({\bf p}) e^{ip\cdot x}+\right .\nonumber\\
&&\left.  \qquad\qquad+\theta (t_1 -t_2) v^1_{\sigma} ({\bf p})
\otimes \overline  v^1_{\sigma} ({\bf p}) e^{-ipx} \right  ] +\\
&+& \left [ \gamma_{\mu\nu} \partial_\mu \partial_\nu + m^2 \right  ]  \int
\frac{d^3 p}{(2\pi)^3 8im^2 E_p}
\left [ \theta (t_2 -t_1) u^2_{\sigma} ({\bf p}) \otimes \overline
u^2_{\sigma} ({\bf p}) e^{ipx}+
\right. \nonumber\\
&&\left. \qquad\qquad+\theta (t_1 -t_2) v^2_{\sigma} ({\bf p})
\otimes \overline  v^2_{\sigma} ({\bf p}) e^{-ipx}\right  ]  +
\mbox{parity-transformed}\,
\sim \delta^{(4)} (x_2 -x_1)\,,\nonumber
\end{eqnarray}
for the bi-vector fields,
see~\cite{dv-hpa} for notation.
The reasons were that the Weinberg equation propagates both causal and tachyonic solutions.} 
The case of the complex-valued vector field will be reported in a separate publication.

%\medskip

{\it Indefinite metrics.}
Usually, one considers the {\it hermitian} field operator in the pseudo-Euclidean netric for the electromagnetic potential:
\begin{equation}
A_\mu = \sum_{\lambda}^{} \int {d^3 {\bf k}\over (2\pi)^3 2E_k} 
[\epsilon_\mu (k,\lambda) a_\lambda ({\bf k}) +\epsilon_\mu^\ast (k,\lambda)
a_\lambda^\dagger ({\bf k})]
\end{equation}
with {\it all} four polarizations to be independent ones. Next, one introduces the Lorentz condition in the weak form
\begin{equation}
[a_{0_t} ({\bf k}) - a_0 ({\bf k})] \vert \phi> =0
\end{equation} 
and the indefinite metrics in the Fock space~\cite[p.90 of the Russian edition]{Bogol}:
$a_{0_t}^\ast = -a_{0_t}$ and $\eta a_\lambda = -a^\lambda \eta$, $\eta^2 =1$,
in order to get the correct sign in the energy-momentum vector
and to not have the problem with the vacuum average.

We observe:
1) that the indefinite metric problems may appear even on the massive level
in the Stueckelberg formalism; 2) The Stueckelberg theory has a good massless limit for propagators, and it reproduces the handling of the indefinite metric in the massless limit (the electromagnetic 4-potential case); 3) we generalized the Stueckelberg formalism (considering, at least, two equations); instead of charge-conjugate solutions we may consider the $P-$  or $CP-$ conjugates. The potential field becomes to be the complex-valued field, that may justify the introduction of the anti-hermitian amplitudes.

In the next section we use the commonly-accepted procedure
for deducing  of higher-spin equations for the case spin-2.

\section{The Standard Formalism. The case of the spin 2.}

We begin with the equations for the 4-rank symmetric spinor:
\begin{eqnarray}
\left [ i\gamma^\mu \partial_\mu - m \right ]_{\alpha\alpha^\prime}
\Psi_{\alpha^\prime \beta\gamma\delta} &=& 0\, ,\\
\left [ i\gamma^\mu \partial_\mu - m \right ]_{\beta\beta^\prime}
\Psi_{\alpha\beta^\prime \gamma\delta} &=& 0\, ,\\
\left [ i\gamma^\mu \partial_\mu - m \right ]_{\gamma\gamma^\prime}
\Psi_{\alpha\beta\gamma^\prime \delta} &=& 0\, ,\\
\left [ i\gamma^\mu \partial_\mu - m \right ]_{\delta\delta^\prime}
\Psi_{\alpha\beta\gamma\delta^\prime} &=& 0\, .
\end{eqnarray} 
The massless limit (if one needs) should be taken in the end of all
calculations.

We proceed expanding the field function in the set of symmetric matrices, 
as in the spin-1 case. In the beginning let us use the
first two indices:\footnote{The matrix $R$ can be related to the
$CP$ operation in the $(1/2,0)\oplus (0,1/2)$ representation.}
\begin{equation} \Psi_{\{\alpha\beta\}\gamma\delta} =
(\gamma_\mu R)_{\alpha\beta} \Psi^\mu_{\gamma\delta}
+(\sigma_{\mu\nu} R)_{\alpha\beta} \Psi^{\mu\nu}_{\gamma\delta}\, .\label{2A}
\end{equation}
We would like to write
the corresponding equations for functions $\Psi^\mu_{\gamma\delta}$
and $\Psi^{\mu\nu}_{\gamma\delta}$ in the form:
\begin{eqnarray}
&&{2\over m} \partial_\mu \Psi^{\mu\nu}_{\gamma\delta} = -
\Psi^\nu_{\gamma\delta}\, , \label{p1}\\
&&\Psi^{\mu\nu}_{\gamma\delta} = {1\over 2m}
\left [ \partial^\mu \Psi^\nu_{\gamma\delta} - \partial^\nu
\Psi^\mu_{\gamma\delta} \right ]\, \label{p2}.
\end{eqnarray} 
Constraints $(1/m) \partial_\mu \Psi^\mu_{\gamma\delta} =0$
and $(1/m) \epsilon^{\mu\nu}_{\quad\alpha\beta}\, \partial_\mu
\Psi^{\alpha\beta}_{\gamma\delta} = 0$ can be regarded as a consequence of
Eqs.  (\ref{p1},\ref{p2}).

Next, we present the vector-spinor and tensor-spinor functions as
\begin{eqnarray}
&&\Psi^\mu_{\{\gamma\delta\}} = (\gamma^\kappa R)_{\gamma\delta}
G_{\kappa}^{\quad \mu} +(\sigma^{\kappa\tau} R )_{\gamma\delta}
F_{\kappa\tau}^{\quad \mu} \, ,\\
&&\Psi^{\mu\nu}_{\{\gamma\delta\}} = (\gamma^\kappa R)_{\gamma\delta}
T_{\kappa}^{\quad \mu\nu} +(\sigma^{\kappa\tau} R )_{\gamma\delta}
R_{\kappa\tau}^{\quad \mu\nu} \, ,
\end{eqnarray}
i.~e.,  using the symmetric matrix coefficients in indices $\gamma$ and
$\delta$. Hence, the total function is
\begin{eqnarray}
\lefteqn{\Psi_{\{\alpha\beta\}\{\gamma\delta\}}
= (\gamma_\mu R)_{\alpha\beta} (\gamma^\kappa R)_{\gamma\delta}
G_\kappa^{\quad \mu} + (\gamma_\mu R)_{\alpha\beta} (\sigma^{\kappa\tau}
R)_{\gamma\delta} F_{\kappa\tau}^{\quad \mu} + } \nonumber\\
&+& (\sigma_{\mu\nu} R)_{\alpha\beta} (\gamma^\kappa R)_{\gamma\delta}
T_\kappa^{\quad \mu\nu} + (\sigma_{\mu\nu} R)_{\alpha\beta}
(\sigma^{\kappa\tau} R)_{\gamma\delta} R_{\kappa\tau}^{\quad\mu\nu};\nonumber\\
\end{eqnarray}
and the resulting tensor equations are:
\begin{eqnarray}
&&{2\over m} \partial_\mu T_\kappa^{\quad \mu\nu} =
-G_{\kappa}^{\quad\nu}\, ,\\
&&{2\over m} \partial_\mu R_{\kappa\tau}^{\quad \mu\nu} =
-F_{\kappa\tau}^{\quad\nu}\, ,\\
&& T_{\kappa}^{\quad \mu\nu} = {1\over 2m} \left [
\partial^\mu G_{\kappa}^{\quad\nu}
- \partial^\nu G_{\kappa}^{\quad \mu} \right ] \, ,\\
&& R_{\kappa\tau}^{\quad \mu\nu} = {1\over 2m} \left [
\partial^\mu F_{\kappa\tau}^{\quad\nu}
- \partial^\nu F_{\kappa\tau}^{\quad \mu} \right ] \, .
\end{eqnarray}
The constraints are re-written to
\begin{eqnarray}
&&{1\over m} \partial_\mu G_\kappa^{\quad\mu} = 0\, ,\quad
{1\over m} \partial_\mu F_{\kappa\tau}^{\quad\mu} =0\, ,\\
&& {1\over m} \epsilon_{\alpha\beta\nu\mu} \partial^\alpha
T_\kappa^{\quad\beta\nu} = 0\, ,\quad
{1\over m} \epsilon_{\alpha\beta\nu\mu} \partial^\alpha
R_{\kappa\tau}^{\quad\beta\nu} = 0\, .
\end{eqnarray}
However, we need to make symmetrization over these two sets
of indices $\{ \alpha\beta \}$ and $\{\gamma\delta \}$. The total
symmetry can be ensured if one contracts the function $\Psi_{\{\alpha\beta
\} \{\gamma \delta \}}$ with {\it antisymmetric} matrices
$R^{-1}_{\beta\gamma}$, $(R^{-1} \gamma^5 )_{\beta\gamma}$ and
$(R^{-1} \gamma^5 \gamma^\lambda )_{\beta\gamma}$ and equate
all these contractions to zero (similar to the $j=3/2$ case
considered in ref.~\cite[p. 44]{Lurie}. We obtain
additional constraints on the tensor field functions:
\begin{eqnarray}
&& G_\mu^{\quad\mu}=0\, , \quad G_{[\kappa \, \mu ]}  = 0\, , \quad
G^{\kappa\mu} = {1\over 2} g^{\kappa\mu} G_\nu^{\quad\nu}\, ,
\label{b1}\\
&&F_{\kappa\mu}^{\quad\mu} = F_{\mu\kappa}^{\quad\mu} = 0\, , \quad
\epsilon^{\kappa\tau\mu\nu} F_{\kappa\tau,\mu} = 0\, ,\\
&& T^{\mu}_{\quad\mu\kappa} =
T^{\mu}_{\quad\kappa\mu} = 0\, ,\quad
\epsilon^{\kappa\tau\mu\nu} T_{\kappa,\tau\mu} = 0\, ,\\
&& F^{\kappa\tau,\mu} = T^{\mu,\kappa\tau}\, ,\quad
\epsilon^{\kappa\tau\mu\lambda} (F_{\kappa\tau,\mu} +
T_{\kappa,\tau\mu})=0\, ,\\
&& R_{\kappa\nu}^{\quad \mu\nu}
= R_{\nu\kappa}^{\quad  \mu\nu} = R_{\kappa\nu}^{\quad\nu\mu}
= R_{\nu\kappa}^{\quad\nu\mu}
= R_{\mu\nu}^{\quad  \mu\nu} = 0\, , \\
&& \epsilon^{\mu\nu\alpha\beta} (g_{\beta\kappa} R_{\mu\tau,
\nu\alpha} - g_{\beta\tau} R_{\nu\alpha,\mu\kappa} ) = 0\, \quad
\epsilon^{\kappa\tau\mu\nu} R_{\kappa\tau,\mu\nu} = 0.\nonumber\\
\label{f1}
\end{eqnarray}
Thus, we  encountered with
the known difficulty of the theory for spin-2 particles in
the Minkowski space.
We explicitly showed that all field functions become to be equal to zero.
Such a situation cannot be considered as a satisfactory one (because it
does not give us any physical information) and can be corrected in several
ways.

\section{The Generalized Formalism.}

We shall modify the formalism in the spirit of  ref.~\cite{Dvo97}.
The field function (\ref{2A}) is now presented as
\begin{equation}
\Psi_{\{\alpha\beta\}\gamma\delta} =
\alpha_1 (\gamma_\mu R)_{\alpha\beta} \Psi^\mu_{\gamma\delta} +
\alpha_2 (\sigma_{\mu\nu} R)_{\alpha\beta} \Psi^{\mu\nu}_{\gamma\delta}
+\alpha_3 (\gamma^5 \sigma_{\mu\nu} R)_{\alpha\beta}
\widetilde \Psi^{\mu\nu}_{\gamma\delta}\, ,
\end{equation}
with
\begin{eqnarray}
&&\Psi^\mu_{\{\gamma\delta\}} = \beta_1 (\gamma^\kappa R)_{\gamma\delta}
G_\kappa^{\quad\mu} + \beta_2 (\sigma^{\kappa\tau} R)_{\gamma\delta}
F_{\kappa\tau}^{\quad\mu} +\beta_3 (\gamma^5 \sigma^{\kappa\tau}
R)_{\gamma\delta} \widetilde F_{\kappa\tau}^{\quad\mu},\nonumber\\
\\
&&\Psi^{\mu\nu}_{\{\gamma\delta\}} =\beta_4 (\gamma^\kappa
R)_{\gamma\delta} T_\kappa^{\quad\mu\nu} + \beta_5 (\sigma^{\kappa\tau}
R)_{\gamma\delta} R_{\kappa\tau}^{\quad\mu\nu} +\beta_6 (\gamma^5
\sigma^{\kappa\tau} R)_{\gamma\delta}
\widetilde R_{\kappa\tau}^{\quad\mu\nu},\nonumber\\
\\
&&\widetilde \Psi^{\mu\nu}_{\{\gamma\delta\}} =\beta_7 (\gamma^\kappa
R)_{\gamma\delta} \widetilde T_\kappa^{\quad\mu\nu} + \beta_8
(\sigma^{\kappa\tau} R)_{\gamma\delta}
\widetilde D_{\kappa\tau}^{\quad\mu\nu}
+\beta_9 (\gamma^5 \sigma^{\kappa\tau} R)_{\gamma\delta}
D_{\kappa\tau}^{\quad\mu\nu}.\nonumber\\
\end{eqnarray}
Hence, the function $\Psi_{\{\alpha\beta\}\{\gamma\delta\}}$
can be expressed as a sum of nine terms:
\begin{eqnarray}
&&\Psi_{\{\alpha\beta\}\{\gamma\delta\}} =
\alpha_1 \beta_1 (\gamma_\mu R)_{\alpha\beta} (\gamma^\kappa
R)_{\gamma\delta} G_\kappa^{\quad\mu} +\alpha_1 \beta_2
(\gamma_\mu R)_{\alpha\beta} (\sigma^{\kappa\tau} R)_{\gamma\delta}
F_{\kappa\tau}^{\quad\mu} + \nonumber\\
&+&\alpha_1 \beta_3 (\gamma_\mu R)_{\alpha\beta}
(\gamma^5 \sigma^{\kappa\tau} R)_{\gamma\delta} \widetilde
F_{\kappa\tau}^{\quad\mu} +
+ \alpha_2 \beta_4 (\sigma_{\mu\nu}
R)_{\alpha\beta} (\gamma^\kappa R)_{\gamma\delta} T_\kappa^{\quad\mu\nu}
+\nonumber\\
&+&\alpha_2 \beta_5 (\sigma_{\mu\nu} R)_{\alpha\beta} (\sigma^{\kappa\tau}
R)_{\gamma\delta} R_{\kappa\tau}^{\quad \mu\nu}
+ \alpha_2
\beta_6 (\sigma_{\mu\nu} R)_{\alpha\beta} (\gamma^5 \sigma^{\kappa\tau}
R)_{\gamma\delta} \widetilde R_{\kappa\tau}^{\quad\mu\nu} +\nonumber\\
&+&\alpha_3 \beta_7 (\gamma^5 \sigma_{\mu\nu} R)_{\alpha\beta}
(\gamma^\kappa R)_{\gamma\delta} \widetilde
T_\kappa^{\quad\mu\nu}+
\alpha_3 \beta_8 (\gamma^5
\sigma_{\mu\nu} R)_{\alpha\beta} (\sigma^{\kappa\tau} R)_{\gamma\delta}
\widetilde D_{\kappa\tau}^{\quad\mu\nu} +\nonumber\\
&+&\alpha_3 \beta_9
(\gamma^5 \sigma_{\mu\nu} R)_{\alpha\beta} (\gamma^5 \sigma^{\kappa\tau}
R)_{\gamma\delta} D_{\kappa\tau}^{\quad \mu\nu}\, .
\label{ffn1}
\end{eqnarray}
The corresponding dynamical
equations are given by the set\footnote{All indices in this formula are
already pure vectorial and have nothing to do with
previous notation. The coefficients $\alpha_i$ and $\beta_i$
may, in general, carry some dimension.}
\begin{eqnarray}
&& {2\alpha_2
\beta_4 \over m} \partial_\nu T_\kappa^{\quad\mu\nu} +{i\alpha_3
\beta_7 \over m} \epsilon^{\mu\nu\alpha\beta} \partial_\nu
\widetilde T_{\kappa,\alpha\beta} = \alpha_1 \beta_1
G_\kappa^{\quad\mu}\,; \label{b}\\
&&{2\alpha_2 \beta_5 \over m} \partial_\nu
R_{\kappa\tau}^{\quad\mu\nu} +{i\alpha_2 \beta_6 \over m}
\epsilon_{\alpha\beta\kappa\tau} \partial_\nu \widetilde R^{\alpha\beta,
\mu\nu} +{i\alpha_3 \beta_8 \over m}
\epsilon^{\mu\nu\alpha\beta}\partial_\nu \widetilde
D_{\kappa\tau,\alpha\beta} - \nonumber\\
&-&{\alpha_3 \beta_9 \over 2}
\epsilon^{\mu\nu\alpha\beta} \epsilon_{\lambda\delta\kappa\tau}
D^{\lambda\delta}_{\quad \alpha\beta} = \alpha_1 \beta_2
F_{\kappa\tau}^{\quad\mu} + {i\alpha_1 \beta_3 \over 2}
\epsilon_{\alpha\beta\kappa\tau} \widetilde F^{\alpha\beta,\mu}\,; \\
&& 2\alpha_2 \beta_4 T_\kappa^{\quad\mu\nu} +i\alpha_3 \beta_7
\epsilon^{\alpha\beta\mu\nu} \widetilde T_{\kappa,\alpha\beta}
=  {\alpha_1 \beta_1 \over m} (\partial^\mu G_\kappa^{\quad \nu}
- \partial^\nu G_\kappa^{\quad\mu})\,; \nonumber\\
\\
&& 2\alpha_2 \beta_5 R_{\kappa\tau}^{\quad\mu\nu} +i\alpha_3 \beta_8
\epsilon^{\alpha\beta\mu\nu} \widetilde D_{\kappa\tau,\alpha\beta}
+i\alpha_2 \beta_6 \epsilon_{\alpha\beta\kappa\tau} \widetilde
R^{\alpha\beta,\mu\nu} -\nonumber\\
&&- {\alpha_3 \beta_9\over 2} \epsilon^{\alpha\beta\mu\nu}
\epsilon_{\lambda\delta\kappa\tau} D^{\lambda\delta}_{\quad \alpha\beta}
= \nonumber\\
&=& {\alpha_1 \beta_2 \over m} (\partial^\mu F_{\kappa\tau}^{\quad \nu}
-\partial^\nu F_{\kappa\tau}^{\quad\mu} ) + {i\alpha_1 \beta_3 \over 2m}
\epsilon_{\alpha\beta\kappa\tau} (\partial^\mu \widetilde
F^{\alpha\beta,\nu} - \partial^\nu \widetilde F^{\alpha\beta,\mu} )\, .\nonumber\\
\label{f}
\end{eqnarray}
Essential constraints are:
\begin{eqnarray}
&&\alpha_1 \beta_1 G^\mu_{\quad\mu} = 0\, ,\quad \alpha_1
\beta_1 G_{[\kappa\mu]} = 0 \, ;  \\
&&\nonumber\\
&&2i\alpha_1 \beta_2 F_{\alpha\mu}^{\quad\mu} +
\alpha_1 \beta_3
\epsilon^{\kappa\tau\mu}_{\quad\alpha} \widetilde F_{\kappa\tau,\mu} =
0\, ;\\
&&\nonumber\\
&&2i\alpha_1 \beta_3 \widetilde F_{\alpha\mu}^{\quad\mu}
+ \alpha_1 \beta_2
\epsilon^{\kappa\tau\mu}_{\quad\alpha} F_{\kappa\tau,\mu} = 0\, ;\\
&&\nonumber\\
&& 2i\alpha_2 \beta_4 T^{\mu}_{\quad\mu\alpha} -
 \alpha_3 \beta_{7}
\epsilon^{\kappa\tau\mu}_{\quad\alpha} \widetilde T_{\kappa,\tau\mu}
= 0\, ;\\
&&\nonumber\\
&& 2i\alpha_3 \beta_{7} \widetilde
T^{\mu}_{\quad\mu\alpha} -
\alpha_2 \beta_4 \epsilon^{\kappa\tau\mu}_{\quad\alpha}
T_{\kappa,\tau\mu} = 0\, ;\\
&&\nonumber\\
&& i\epsilon^{\mu\nu\kappa\tau} \left [ \alpha_2 \beta_6 \widetilde
R_{\kappa\tau,\mu\nu} + \alpha_3 \beta_{8} \widetilde
D_{\kappa\tau,\mu\nu} \right ] + 2\alpha_2 \beta_5
R^{\mu\nu}_{\quad\mu\nu}  + 2\alpha_3
\beta_{9} D^{\mu\nu}_{\quad \mu\nu}  = 0\, ;\\
&&\nonumber\\
&& i\epsilon^{\mu\nu\kappa\tau} \left [ \alpha_2 \beta_5 R_{\kappa\tau,
\mu\nu} + \alpha_3 \beta_{9} D_{\kappa\tau, \mu\nu} \right ]
+ 2\alpha_2 \beta_6 \widetilde R^{\mu\nu}_{\quad\mu\nu}
+ 2\alpha_3 \beta_{8} \widetilde D^{\mu\nu}_{\quad\mu\nu}  =0\, ;\\
&&\nonumber\\
&& 2i \alpha_2 \beta_5 R_{\beta\mu}^{\quad\mu\alpha} + 2i\alpha_3
\beta_{9} D_{\beta\mu}^{\quad\mu\alpha} + \alpha_2 \beta_6
\epsilon^{\nu\alpha}_{\quad\lambda\beta} \widetilde
R^{\lambda\mu}_{\quad\mu\nu} +\alpha_3 \beta_{8}
\epsilon^{\nu\alpha}_{\quad\lambda\beta} \widetilde
D^{\lambda\mu}_{\quad \mu\nu} = 0\, ;\nonumber\\
&&\\
&&2i\alpha_1 \beta_2 F^{\lambda\mu}_{\quad\mu} - 2 i \alpha_2 \beta_4
T_\mu^{\quad\mu\lambda} + \alpha_1 \beta_3 \epsilon^{\kappa\tau\mu\lambda}
\widetilde F_{\kappa\tau,\mu} +\alpha_3 \beta_7
\epsilon^{\kappa\tau\mu\lambda} \widetilde T_{\kappa,\tau\mu} =0\, ;\\
&&\nonumber\\
&&2i\alpha_1 \beta_3 \widetilde F^{\lambda\mu}_{\quad\mu} - 2 i \alpha_3
\beta_7 \widetilde T_\mu^{\quad\mu\lambda} + \alpha_1 \beta_2
\epsilon^{\kappa\tau\mu\lambda} F_{\kappa\tau,\mu} +\alpha_2
\beta_4 \epsilon^{\kappa\tau\mu\lambda}  T_{\kappa,\tau\mu} =0\, ;\\
&&\nonumber\\
&&\alpha_1 \beta_1 (2G^\lambda_{\quad\alpha} - g^\lambda_{\quad\alpha}
G^\mu_{\quad\mu} ) - 2\alpha_2 \beta_5 (2R^{\lambda\mu}_{\quad\mu\alpha}
+2R_{\alpha\mu}^{\quad\mu\lambda} + g^\lambda_{\quad\alpha}
R^{\mu\nu}_{\quad\mu\nu}) +\nonumber\\
&+& 2\alpha_3 \beta_9
(2D^{\lambda\mu}_{\quad\mu\alpha} + 2D_{\alpha\mu}^{\quad\mu\lambda}
+g^\lambda_{\quad\alpha} D^{\mu\nu}_{\quad\mu\nu})+\nonumber\\
&+&2i\alpha_3 \beta_8 (\epsilon_{\kappa\alpha}^{\quad\mu\nu}
\widetilde D^{\kappa\lambda}_{\quad\mu\nu} -
\epsilon^{\kappa\tau\mu\lambda} \widetilde D_{\kappa\tau,\mu\alpha}) -
\nonumber\\
&-& 2i\alpha_2 \beta_6 (\epsilon_{\kappa\alpha}^{\quad \mu\nu}
\widetilde R^{\kappa\lambda}_{\quad\mu\nu} -
\epsilon^{\kappa\tau\mu\lambda} \widetilde R_{\kappa\tau,\mu\alpha})
= 0\, ; \\
&&\nonumber\\
&& 2\alpha_3 \beta_8 (2\widetilde D^{\lambda\mu}_{\quad\mu\alpha} + 2
\widetilde D_{\alpha\mu}^{\quad\mu\lambda} +g^\lambda_{\quad\alpha}
\widetilde D^{\mu\nu}_{\quad\mu\nu}) - 2\alpha_2 \beta_6 (2\widetilde
R^{\lambda\mu}_{\quad\mu\alpha} +2 \widetilde
R_{\alpha\mu}^{\quad\mu\lambda} + \nonumber\\
&+& g^\lambda_{\quad\alpha} \widetilde
R^{\mu\nu}_{\quad\mu\nu}) +
+ 2i\alpha_3 \beta_9 (\epsilon_{\kappa\alpha}^{\quad\mu\nu}
D^{\kappa\lambda}_{\quad\mu\nu}  - \epsilon^{\kappa\tau\mu\lambda}
D_{\kappa\tau,\mu\alpha} ) -\nonumber\\
&-& 2i\alpha_2 \beta_5
(\epsilon_{\kappa\alpha}^{\quad\mu\nu} R^{\kappa\lambda}_{\quad\mu\nu}
- \epsilon^{\kappa\tau\mu\lambda} R_{\kappa\tau,\mu\alpha} ) =0\, ;\\
&&\nonumber\\
&&\alpha_1 \beta_2 (F^{\alpha\beta,\lambda} - 2F^{\beta\lambda,\alpha}
+ F^{\beta\mu}_{\quad\mu}\, g^{\lambda\alpha} - F^{\alpha\mu}_{\quad\mu}
\, g^{\lambda\beta} ) - \nonumber\\
&-&\alpha_2 \beta_4 (T^{\lambda,\alpha\beta}
-2T^{\beta,\lambda\alpha} + T_\mu^{\quad\mu\alpha} g^{\lambda\beta} -
T_\mu^{\quad\mu\beta} g^{\lambda\alpha} ) +\nonumber\\
&+&{i\over 2} \alpha_1 \beta_3 (\epsilon^{\kappa\tau\alpha\beta}
\widetilde F_{\kappa\tau}^{\quad\lambda} +
2\epsilon^{\lambda\kappa\alpha\beta} \widetilde F_{\kappa\mu}^{\quad\mu} +
2 \epsilon^{\mu\kappa\alpha\beta} \widetilde F^\lambda_{\quad\kappa,\mu})
-\nonumber\\
&-& {i\over 2} \alpha_3 \beta_7 ( \epsilon^{\mu\nu\alpha\beta} \widetilde
T^{\lambda}_{\quad\mu\nu} +2 \epsilon^{\nu\lambda\alpha\beta} \widetilde
T^\mu_{\quad\mu\nu} +2 \epsilon^{\mu\kappa\alpha\beta} \widetilde
T_{\kappa,\mu}^{\quad\lambda} ) =0\, .
\end{eqnarray}
They are  the results of contractions of the field function (\ref{ffn1})
with three antisymmetric matrices, as above. Furthermore,
one should recover the relations (\ref{b1}-\ref{f1}) in the particular
case when $\alpha_3 = \beta_3 =\beta_6 = \beta_9 = 0$ and
$\alpha_1 = \alpha_2 = \beta_1 =\beta_2 =\beta_4
=\beta_5 = \beta_7 =\beta_8 =1$.

As a discussion we note that in such a framework we already have physical
content because only certain combinations of field functions
would be equal to zero. In general, the fields
$F_{\kappa\tau}^{\quad\mu}$, $\widetilde F_{\kappa\tau}^{\quad\mu}$,
$T_{\kappa}^{\quad\mu\nu}$, $\widetilde T_{\kappa}^{\quad\mu\nu}$, and
$R_{\kappa\tau}^{\quad\mu\nu}$,  $\widetilde
R_{\kappa\tau}^{\quad\mu\nu}$, $D_{\kappa\tau}^{\quad\mu\nu}$, $\widetilde
D_{\kappa\tau}^{\quad\mu\nu}$ can  correspond to different physical states
and the equations above describe oscillations one state to another.

Furthermore, from the set of equations (\ref{b}-\ref{f}) one
obtains the {\it second}-order equation for symmetric traceless tensor of
the second rank ($\alpha_1 \neq 0$, $\beta_1 \neq 0$):
\begin{equation} {1\over m^2} \left [\partial_\nu
\partial^\mu G_\kappa^{\quad \nu} - \partial_\nu \partial^\nu
G_\kappa^{\quad\mu} \right ] =  G_\kappa^{\quad \mu}\, .
\end{equation}
After the contraction in indices $\kappa$ and $\mu$ this equation is
reduced to the set
\begin{eqnarray}
&&\partial_\mu G_{\quad\kappa}^{\mu} = F_\kappa\,,  \\
&&{1\over m^2} \partial_\kappa F^\kappa = 0\, ,
\end{eqnarray}
i.~e.,  to the equations connecting the analogue of the energy-momentum
tensor and the analogue of the 4-vector potential. As we showed in our
recent work~\cite{Dvo97} the longitudinal potential is perfectly suitable
for construction of electromagnetism (see also recent works on the notoph
and notivarg concept~\cite{Tybor}).  Moreover, according to the Weinberg
theorem~\cite{WTH} for massless particles it is the gauge part of the
4-vector potential which is the physical field. The case, when the
longitudinal potential is equated to zero, can be considered as a
particular case only.  This case may be relevant to some physical
situation but hardly to be considered as a fundamental one.

\section{Conclusions}

\begin{itemize}

\item
The $(1/2,1/2)$ representation contains both the  spin-1 and spin-0
states (cf. with the Stueckelberg formalism).

\item
Unless we take into account the fourth state (the ``time-like" state, or
the spin-0 state) the set of 4-vectors is {\it not} a complete set in a mathematical sense.

\item
We cannot remove terms like $(\partial_\mu B^\ast_\mu)(\partial_\nu B_\nu)$ 
terms from the Lagrangian and dynamical invariants unless apply the Fermi 
method, i.~e., manually. The Lorentz condition applies only to the spin 1 states.

\item
We have some additional terms in the expressions of the energy-mo\-men\-tum vector (and, accordingly, of the 4-current and the Pauli-Lunbanski vectors), which are the consequence of the impossibility to apply the Lorentz condition for spin-0 states.

\item
Helicity vectors are not eigenvectors of the parity operator. Meanwhile, the parity is a ``good" quantum number, $[{\cal P}, {\cal H}]_- =0$ in the Fock space.

\item
We are able to describe the states of different masses in this representation from the beginning.

\item
Various-type field operators can be constructed in the $(1/2,1/2)$ representation space. For instance, they can contain $C$, $P$ and $CP$ conjugate states.
Even if $b_\lambda^\dagger =a_\lambda^\dagger$ 
we can have complex 4-vector fields.
We found the relations between creation, annihilation operators for different types of the field operators $B_\mu$.

\item
Propagators have good behavious in the massless limit as opposed to those of the Proca theory.

\item
The spin-2 case can be considered on an equal footing with the spin-1 case. Further investigations may provide additional foundations to ``surprising" similarities of gravitational and electromagnetic
equations in the low-velocity limit, refs.~\cite{Wein2,Jef}.

\end{itemize}

\end{document}